\documentclass[12pt]{article}
\usepackage[utf8]{inputenc}
\usepackage{gensymb}
\usepackage{geometry}
\usepackage[flushmargin]{footmisc}
\usepackage{setspace}
\usepackage{graphicx}
\usepackage{amsfonts}
\usepackage{mathtools}
\usepackage{amsmath}
\graphicspath{ {./figures/} }
\usepackage{mathpazo}
\usepackage{todonotes}
\usepackage{pdfpages}
\usepackage{natbib}
\usepackage{enumitem}
\usepackage{hyperref}
\hypersetup{
    colorlinks=true,
    urlcolor=blue,
    linkcolor=black,
    citecolor=black,
}

\newcounter{claims}
\newenvironment{claim}{\begin{quote} \refstepcounter{claims} \textbf{Claim \theclaims:} }{\end{quote}}

\DeclareMathOperator*{\argmax}{arg\,max}

\title{Industrial Policy for Advanced AI: \\ Compute Pricing and the Safety Tax}

\author{Mckay Jensen, Nicholas Emery-Xu, Robert Trager}
\date{\today}

\begin{document}
    
\maketitle

\begin{abstract}
    
    \onehalfspacing
    Using a model in which agents compete to develop a potentially dangerous new technology (AI), we study how changes in the pricing of factors of production (computational resources) affect agents' strategies, particularly their spending on safety meant to reduce the danger from the new technology. In the model, agents split spending between safety and performance, with safety determining the probability of a ``disaster" outcome, and performance determining the agents' competitiveness relative to their peers. For given parameterizations, we determine the theoretically optimal spending strategies by numerically computing Nash equilibria. Using this approach we find that (1) in symmetric scenarios, compute price increases are safety-promoting if and only if the production of performance scales faster than the production of safety; (2) the probability of a disaster can be made arbitrarily low by providing a sufficiently large subsidy to a single agent; (3) when agents differ in productivity, providing a subsidy to the more productive agent is often better for aggregate safety than providing the same subsidy to other agent(s) (with some qualifications, which we discuss);  (4) when one agent is \textit{much more} safety-conscious, in the sense of believing that safety is more difficult to achieve, relative to his competitors, subsidizing that agent is typically better for aggregate safety than subsidizing its competitors; however, subsidizing an agent that is only \textit{somewhat} more safety-conscious often decreases safety. Thus, although subsidizing a much more safety-conscious, or productive, agent often improves safety as intuition suggests, subsidizing a somewhat more safety-conscious or productive agent can often be harmful.
\end{abstract}

\newpage
\section{Introduction}

Rapid advances in artificial intelligence (AI) systems have led to concerns about the alignment of such systems with human values, especially as they come to influence decision-making over increasingly significant aspects of human society (\cite{christiano_alignment, yudkowsky_intelligence_2013}). These risks are exacerbated by the strategic environment in which AI developers find themselves. If misalignment risks are not fully internalized by developers, they may have an incentive to reduce safety investments in favor of investments in performance to increase the chance of being the first to develop new technologies. Such a \textbf{safety-performance tradeoff} (\cite{trager_strategic_2021}) is an example of what \cite{christiano_alignment} calls a \textbf{safety tax}, or the marginal cost of deploying an AI system aligned with human values over an equivalent but unaligned system.

As a result, within the AI governance field, the development of mechanisms to reduce safety taxes is an active area of research. In the present work, we study the role of input pricing in reducing the risk from such a competitive scenario. We develop a formal model in which agents are racing to develop a novel AI system. Each agent purchases an input, computation, which is allocated between investments in performance or safety. agents' relative levels of performance determine the probabilities of each agent winning the race, while safety investments reduce the risk of a disaster that negatively impacts all players. Formalizing the tradeoff between safety and performance in this way allows us to study how agents respond to changes in the prices of factors of production, which is a key contribution of this work; in the context of AI technology, our model allows us to consider how changes in compute prices are likely to affect safety. We consider the problem of a principal who wants to increase equilibrium safety and is able to influence compute pricing (or more generally, the price of some key production input) to that end. We investigate how compute price changes for one or multiple agents affect equilibrium safety.

We solve computationally for the Nash equilibrium levels of safety and performance investments and derive four main results.

First, restricting the principal to set a single price for all agents, we show that safety is increasing in compute price if and only if the elasticity of safety with respect to spending on safety is greater than the elasticity of safety with respect to spending on performance. In this case, increasing the input cost is beneficial for safety because agents will reduce performance more than safety for a given increase in price. This result implies that safety declines as the price of compute declines if the effort required to make systems safe increases enough in the performance of the system.

Second, we allow the principal to set individual prices for agents. This might be accomplished, for example, by setting a single price for renting cloud compute and then offering differential subsidies to firms.\footnote{This is a regular practice for providers of cloud compute. See, for example, OpenAI's partnership with Microsoft Azure or C3.ai's partnership with Google Cloud.} We find that arbitrarily high safety can be achieved in equilibrium by providing a sufficient subsidy to a single agent. By giving one agent a large enough advantage in the race, the subsidized agent can afford to both devote sufficient resources to win the race and produce a high level of safety. 

Third, if one agent is more efficient at producing performance, we find that the principals should subsidize the more productive agent when the performance elasticity of safety is high. When the elasticity is low -- for instance, when increasing performance decreases the effort required to make a system safe -- subsidizing the less productive agent reduces the equilibrium difference in capabilities of the agents. In this case, the developers face little or no safety-performance tradeoff, and thus they continue to maintain high levels of safety, even when they have similar levels of capabilities. In what is probably the more likely case, when the safety elasticity of performance is high (implying a significant safety-performance tradeoff), providing subsidies to bring agents' capabilities closer together is not beneficial in this model because they are then incentivized to race to the bottom by cutting corners on safety. In this latter case, a principal should instead subsidize the more productive agent, increasing her probability of winning and allowing her to choose a higher level of safety.

Fourth, we examine scenarios in which agents differ in their attitudes toward the risk of a disaster. In particular, we find that, given some reasonable assumptions, when agents have sufficiently different beliefs about the cost of achieving a given level of safety, providing a subsidy to an agent who believes safety to be costly to achieve is better for aggregate safety than providing the same subsidy to an agent who believes safety to be relatively easy to achieve. This matches the intuition that assisting safety-conscious agents is better for safety than assisting their competitors; however, there are some cases under which this intuition fails, so we also examine some of those cases. In particular, subsidies for safety-conscious agents are not reliably safety-promoting when the differences in safety-consciousness between them and their competitors are not large or if we use some other definition of safety-consciousness.

\section{Risk and compute pricing}

\subsection{Mechanisms for reducing risk}

Following the work by \cite{armstrong_racing_2016}, a growing body of literature has sought to understand how the strategic environment in a technology race affects risk and uncover mechanisms to reduce it. Factors that have been identified as being important to risk are agents' knowledge about capabilities (\cite{emery-xu_uncertainty_2022, armstrong_racing_2016}), the capability gap between the leader and her competitors (\cite{stafford_IAEA_2022, stafford_international_2021}), and the influence of safety on both development speed and the probability of risk (\cite{han_mediating_2021}).

A variety of mechanisms have been proposed to reduce these risks. \cite{han_mediating_2021} consider the conditions under which a government can use taxes to punish unsafe development or subsidies to reward safe development, finding that both interventions reduce risk under certain conditions but only taxes can lead to overregulation and a suboptimal reduction in innovative output. This work assumes that AI development is proceeding without international cooperation or competion. The global nature of contemporary AI development, however, hinders the efficacy of government regulation, as countries may have an incentive to underprovide regulation in order to outcompete their rivals. Other scholars have, therefore, focused on mechanisms to which agents will voluntarily agree. Drawing from the success of the Nuclear Non-Proliferation Treaty, \cite{stafford_IAEA_2022} study the role of information sharing agreements in reducing risk, finding that if agents are not too close in capability, the leader has an incentive to share some technology with the laggard in return for the latter exiting the race. \cite{emery-xu_uncertainty_2022} find that, except when the race is highly rivalrous and cutting corners on safety can approximately guarantee that a agent wins the race, public revelation of capabilities reduces risk.

While all of these models assume that agents' capabilities are exogenously endowed by nature, AI developers must purchase research performance in competitive markets for human capital, computational capital, and other inputs (\cite{noauthor_why_nodate, noauthor_ai_nodate}). Thus, an input producer with market power has the ability to influence the safety choices of agents. Governments, through industrial policy, and other agents, through technical collaborations and subsidies, can influence equilibrium risk levels. Because our baseline model analyzes a complete information scenario, we can allow the principal to implement first-degree price discrimination and thus study the effects of the first-best pricing strategy. However, even though the principal can observe the agents' types, we still observe a moral hazard constraint stemming from the dual-use nature of compute.

\subsection{Compute scaling}

Our present work simplifies the production process by focusing on a single input - computation. Why? First, we do so to simplify the analysis and make the results easier to interpret. We encourage future researchers to build on these results in considering other inputs into production functions. Second, computational capital plays a key role in driving progress in deep learning, the most prominent AI paradigm, compared to other US R\&D sectors (\cite{besiroglu_economic_2022}). Third, because physical capital is more accumulable following a change in price than is labor, it is relatively easier for agents to respond to a price change in computation than to a wage change for researchers. 

How, then, does computation translate into AI progress? We assume it takes the following power-law form:
\begin{align}
    p := BX_p^\beta
\end{align}
where $X_p$ is the amount of compute used to advance the performance of the system, $p$ is the level of performance and the other parameters are constants of the production function. We use this functional form because experimental results have shown that neural network performance tends to scale in this way with respect to computation. (\cite{jones2021scaling, henighan2020scaling, kaplan2020scaling, lepikhin2020gshard, hestness2017deep}).\footnote{Fortuitously, by assuming a power-law relationship, we can consider our model a special case of the canonical ``ideas production function" in endogenous growth theory (\cite{jones_r_1995, romer_endogenous_1990}), which takes the form $p\equiv\frac{\Dot{A}}{A} = A^{\nu-1}K_p^\beta$. With $\nu=1$, we recover our model.} \cite{thompson_computational_2020} shows that, across a wide variety of machine learning benchmarks, performance is highly dependent upon the level of computational inputs.\footnote{\citet{hoffman2022} show that other inputs, in particular the size of the training dataset, are also important - scaling compute without scaling data is not efficient.} We also assume safety research follows a similar scaling law. While there exists far less research on how safety scales with computation, there exists evidence that safety outputs scale with compute according to a power law on some AI safety benchmarks (\cite{bai_training_2022, askell_general_2021}).

\section{The model}

There are $n$ players, and each player $i = 1, 2, \dots, n$ chooses, simultaneously, to purchase some amount $X_i$ of a factor of production (compute power), at a per-unit price $r$, and divides it between creating performance and creating safety. Thus, $X_i = X_{s,i} + X_{p,i}$, where $X_{s,i}$ and $X_{p,i}$ are the amounts of the factor of production used for safety and performance, respectively. Safety ($s$) and performance ($p$) are produced according to the following production functions:
\begin{equation}\label{eq:def-s}
    s_i := A_i X_{s,i}^{\alpha_i} p_i^{-\theta_i}
\end{equation}
\begin{equation}\label{eq:def-p}
    p_i := B_i X_{p,i}^{\beta_i}
\end{equation}
Note that $p_i$ appears in equation \eqref{eq:def-s} to reflect the idea that safety may become more expensive as performance increases, corresponding to the case where $\theta_i > 0$. $\alpha_i$ is the compute elasticity of safety, describing how well safety progress scales with additional compute dedicated to safety. $\beta_i$ is the analogous parameter for performance research. Finally, $\theta_i$ controls the degree of the safety-performance tradeoff. In particular, $-\theta_i$ is the elasticity of safety with respect to performance (the $p_i$-elasticity of $s_i$); when we let $\theta_i > 0$, spending on performance has a negative impact on safety. A high value of $\theta_i$ indicates that there is a large safety tax, as there is a large cost to performance in investing in safe systems, while a low or even negative value of $\theta_i$ indicates that the safety tax is small or nonexistent. This might be the case when only safe systems perform well as evaluated by the market. For example, consumers are unlikely to purchase autonomous vehicles that do not exhibit a high degree of both performance and safety.\footnote{There is an interesting distinction between performance driven by investment, which we analyze here, and performance level driven by implementation choices. An example of the latter is enabling a self-driving car mode to operate only on highways or also on city streets. In such cases, we expect higher performance to imply a higher cost for an equivalent levels of safety - a safety tax. See \cite{trager_strategic_2021}.}

\subsection{Payoffs and players' objectives}

Players (agents) compete in a contest, where the probability that player $i$ wins is defined as the simple contest success function
\begin{equation}\label{eq:def-q}
    q_i := \frac{p_i}{\sum_{j=1}^n p_j}.
\end{equation}

At the same time, players' realized levels of safety aggregate to produce some probability that a disaster occurs (discussed more in section \ref{subsec:disaster-risk-agg}); we define $\sigma_i$ as the probability of a safe outcome (no disaster), given that player $i$ wins the contest. If player $i$ wins the contest, and none of the players cause a disaster, player $i$ gets a normalized payoff of $1$. Players that do not win receive a payoff of $0$. If there is a disaster, none of the players get a payoff for winning the contest; instead, each player pays a disaster cost $d_i$.

Putting this all together, player $i$'s expected net payoff is
\begin{equation}\label{eq:def-payoff}
    u_i := \sigma_i q_i - \left(1 - \sum_j \sigma_j q_j \right) d_i - r(X_{s,i} + X_{p,i}).
\end{equation}

\subsection{Disaster risk aggregation}\label{subsec:disaster-risk-agg}

The players' realized level of safety determines the probability that a disaster occurs. Here, we focus on two different ways of aggregating player safety choices to determine that probability.

\subsubsection{Independent (multiplicative) disaster risks}

In our base case, each player has some independent probability of causing a disaster; $s_i$ represents the odds that player $i$ \textit{does not} cause a disaster. Thus, $s_i / (1 + s_i)$ is the probability that player $i$ does not cause a disaster, and
\begin{equation}\label{eq:def-sigma}
    \sigma := \prod_{j=1}^n \frac{s_j}{1+s_j}
\end{equation}
is the probability that none of the players cause a disaster. (We refer to $\sigma$ as the ``aggregate safety.") Note that this probability is the same regardless of who wins the contest; i.e., $\sigma_i = \sigma$, and thus we can simplify equation \eqref{eq:def-payoff} to
\begin{equation}
    u_i = \sigma q_i - (1 - \sigma) d_i - r(X_{s,i} + X_{p,i})
\end{equation}
in this case.

\subsubsection{Disaster risk only from contest winner}

As an alternate case, we can assume that instead of each player carrying some independent risk of causing a disaster, only the winner of the contest can cause a disaster. That is, we make the assumption that $s_i / (1+s_i)$ is the probability of a safe outcome, conditional on player $i$ being the contest winner:
\begin{equation}
    \sigma_i = \frac{s_i}{1 + s_i}
\end{equation}

The aggregate safety in this case (unconditional probability that no player causes a disaster) is
\begin{equation}
    \sigma = \sum_{i = 1}^n \sigma_i q_i = \sum_{i=1}^n \left(\frac{s_i}{1+s_i}\right) q_i.
\end{equation}

\subsection{Heterogeneous beliefs}

Up to this point, we've assume that all players have the same (correct) beliefs about the model parameters. Unless stated otherwise, this will be our default assumption, but we can also consider cases where players disagree on those parameters' values. In this case, we assume that each player $i$'s objective is to maximize $u_i$ subject to their own beliefs about the model parameters; we also assume that all players are accurately informed of each other's beliefs (i.e., higher-order beliefs are perfect).

In this paper, we will be particularly interested in heterogenous beliefs about the parameter $A$ (the safety productivity factor). When players have different beliefs about $A$, they have different beliefs about the cost of achieving a given level of safety: a player that believes that $A$ is higher believes safety is cheap and therefore that less investment is required to reduce the risk of a disaster. Thus players' beliefs about $A$ can be used as a measure of their safety-consciousness -- we say that players who believe $A$ to be low (i.e. believe safety to be expensive) are safety-conscious, and conversely for players who believe $A$ to be high.

\subsection{Solution criterion}

We look for pure-strategy Nash equilibria for this model.\footnote{Although mixed-strategy equilibria or multiple pure-strategy equilibria may exist for a given parameterization of the model, we have found that this is rarely the case unless there is some discontinuity in payoffs based on player strategies. One situation in which this may occur is if we choose extreme parameter values that result in situations where players may sometimes prefer not to produce at all. However, studying such scenarios is not the focus of this paper, since the version of the model we use here does not allow players to enter or fully exit competition and therefore is likely to reflect these scenarios poorly. Extending our model to allow for player entry/exit may be a worthwhile way to expand on this work.} That is, we find solutions where each player $i$ chooses $X_{s,i}$, $X_{p,i}$ such that $u_i$ is optimal, given the other players' choices. Due to the intractability of finding a closed-form solution, we implement a computational approach to solve for equilibrium values of $X_{s,i}$, $X_{p,i}$. A description of our solver is presented in Appendix \ref{sec:appendix-solver-description}.

\section{Response of safety to changes in input cost}

In this section, we analyze how the principal can use the input price $r$ to affect the probability of a safe outcome. For simplicity, we analyze the case with two players and begin by assuming the principal can only set a single price. The first two claims presented here are true for both risk aggregation assumptions presented in section \ref{subsec:disaster-risk-agg}, while the later claims are sensitive to that assumption.

\begin{claim}\label{claim:crit-theta}
When players are identical, the probability \(\sigma\) of a safe outcome increases with the factor price \(r\) if and only if $\theta > \alpha / \beta$.
\end{claim}

\begin{figure}[h]
    \centering
    \includegraphics[width=0.8\textwidth]{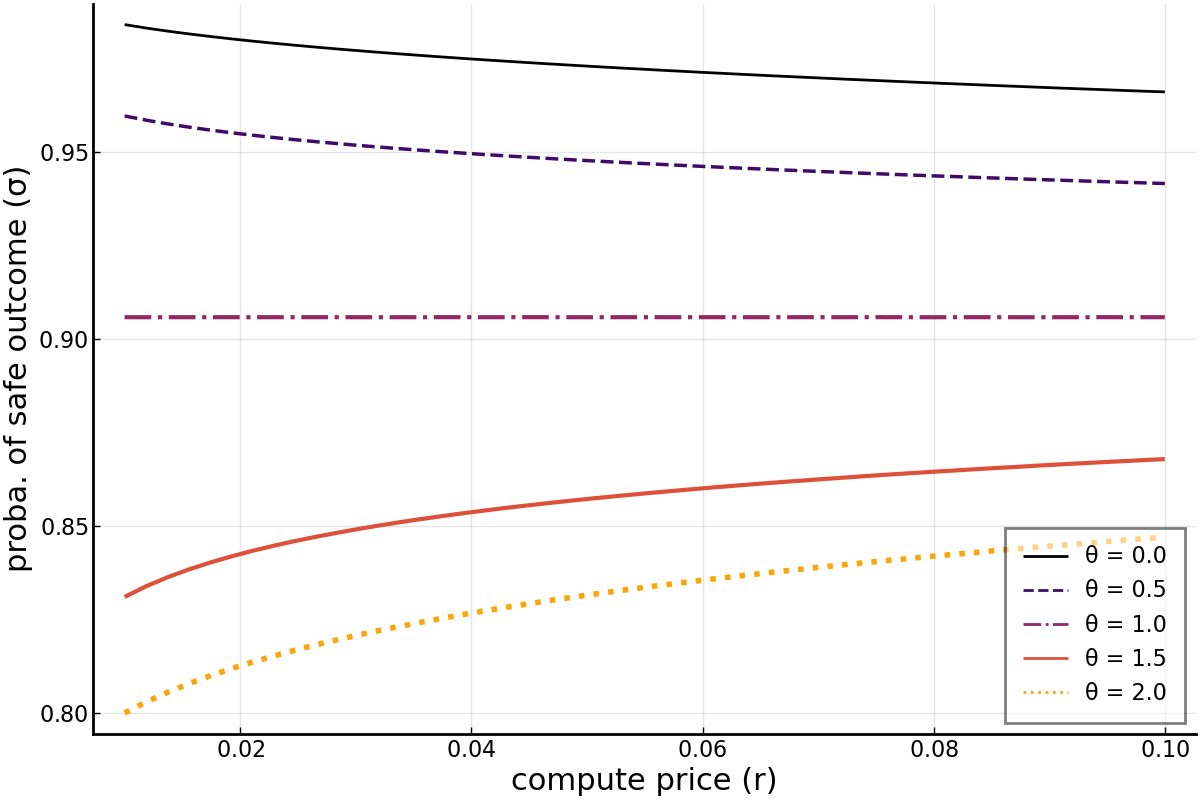}
    \caption{\textbf{The effect of price changes on safety depends on the scaling of the safety-performance tradeoff.} Here, \(\alpha = \beta = 0.5\). Aggregate safety \(\sigma\) increases with \(r\) iff \(\theta > 1 = \alpha / \beta\).}
    \label{fig:claim-theta-regimes}
\end{figure}

A typical scenario illustrating this claim is shown in Figure \ref{fig:claim-theta-regimes}. In the figure, $\alpha=\beta=0.5$ so that safety and performance each scale with the square root of compute. Thus, safety increases in the price of compute for $\theta > 1$ and decreases for $\theta < 1$.

Suppose that we start at some level of inputs, $X_s$ and $X_p$, and scale both up by a factor of $c > 1$. We have
\begin{equation}
    s(cX_s, cX_p) = c^{\alpha - \theta\beta} \frac{A}{B^\theta} X_s^\alpha X_p^{-\theta \beta} = c^{\alpha - \theta\beta} s(X_s, X_p),
\end{equation}
meaning that this scaling-up of inputs results in increased safety if and only if $\alpha > \theta \beta$. For $\alpha \leq \theta \beta$, if we want to increase safety, we must increase $X_s$ at a greater rate than we increase $X_p$.

We can also see this by combining equations \eqref{eq:def-s} and \eqref{eq:def-p}, giving us
\begin{equation}
    s = \frac{A}{B^\theta} X_s^\alpha X_p^{-\theta \beta},
\end{equation}
meaning that the elasticity of safety with respect to $X_p$ is $-\theta \beta$. Recalling that $\alpha$ represents the elasticity of safety with respect to $X_s$, we can see that the elasticity of safety with respect to a uniform increase in $X_s$ and $X_p$ is $\alpha - \theta \beta$. Safety's returns to scale are determined by the sign of this quantity.

We can thus interpret Claim 1 as saying that \(\sigma\) increases with \(r\) if and only if safety has negative returns to scale in all its inputs (i.e., if safety is decreases when all outputs are scaled up uniformly). More loosely, we can think of this as saying that price increases are safety-promoting when production of performance outpaces production of safety.

We now consider outcomes in which the principal can engage in first-degree price discrimination and charge $r_i$ based on observed characteristics of the agents.

\begin{claim}\label{claim:decisive-advantage} When $d > 0$, arbitrarily high probabilities of a safe outcome can be achieved by giving a single player (and not that player's competitors) a sufficiently low factor price.\end{claim}

\begin{figure}
    \centering
    \includegraphics[width=0.8\textwidth]{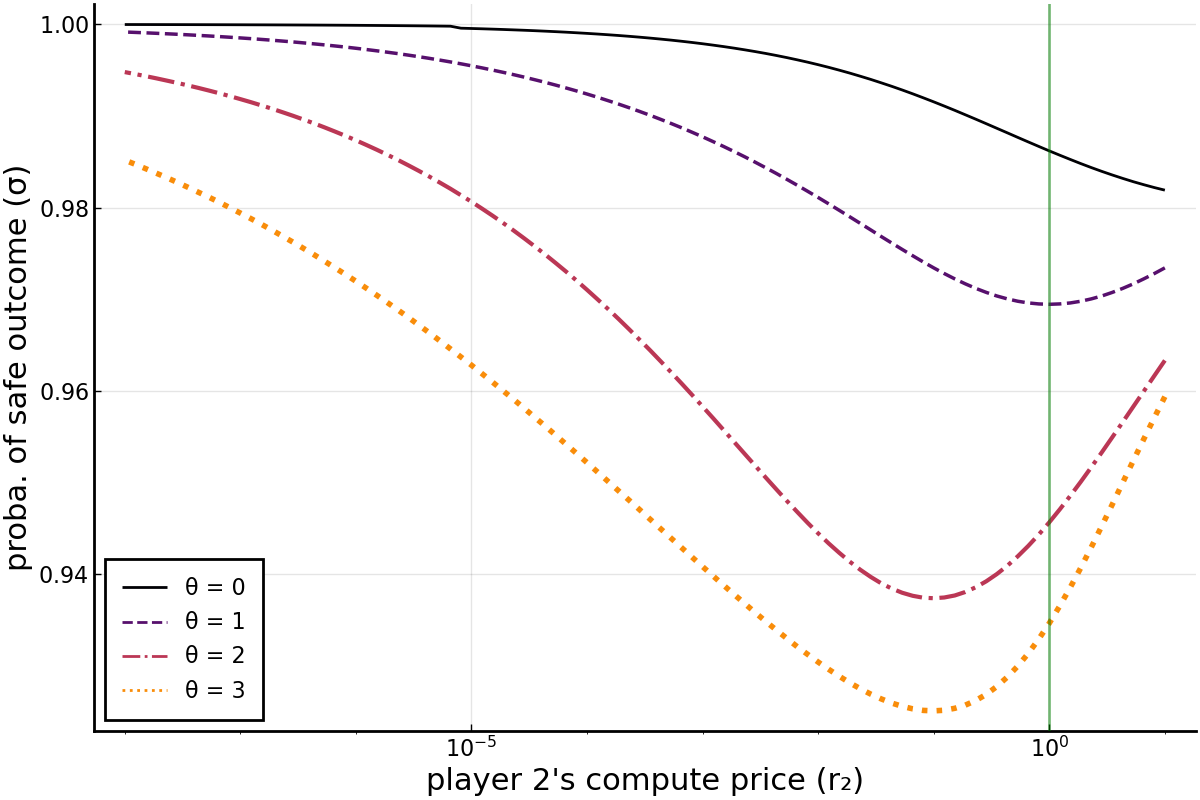}
    \caption{\textbf{Probability of a safe outcome goes to 1 as a single player's compute price goes to 0.} Aggregate safety \(\sigma\) is shown as player 2's \(r\) varies and player 1's \(r\) is held fixed at \(r_1 = 1\). (Dashed line marks \(r_1\).) In all cases, probability of a safe outcome ($\sigma$) converges to 1 as $r_2 \rightarrow 0$.}
    \label{fig:claim-large-subsidies}
\end{figure}

A typical scenario illustrating this claim is shown in Figure \ref{fig:claim-large-subsidies}. Player 1's price of compute is held constant while player 2's is allowed to vary. Moving from right to left, we see that decreasing the price of compute at first decreases safety when $\theta$ is not too low. This happens for essentially the same reasons discussed in relation to Claim 1. As player 2's price gets even lower, however, the probability of a safe outcome goes to 1. 

Thus, in symmetric cases where players have the same factor price $r$, the relationship between that price and safety is as stated in Claim 1. In general, giving a single player a subsidy (lower $r$) is not necessarily safety-promoting, and in asymmetric cases, giving a subsidy to different players can have different effects on safety. The intuition for Claim 2 is that, although small subsidies to a single player may not increase safety, if we give a player enough of a subsidy that all other players become practically unable to compete, then the subsidized player is able to take their focus off of performance and use their inexpensive resources to achieve a high level of safety.\footnote{An important caveat is that this is dependent on the assumption that only the potential reward for winning is fixed, with players' relative levels of performance determining who wins the contest but not the size of the reward for winning. If performance has some intrinsic benefit (e.g., if we think that players value creating advanced AI sooner, even in the absence of competition), then this claim will not be strictly true.}

The next claim examines the case where one player is more effective at converting resources into performance than their competitor. In this case, whether risks come all players, or only from the winner of the technology competition, becomes significant for the findings. 

\begin{claim}\label{claim:productivity-advantage}
    Suppose that one player is more productive at producing performance (has a higher \(B\)) relative to their competitor(s), with players otherwise
    identical.
    
    \begin{enumerate}[label = \textbf{(3.\alph*)}]
        \item In the case of multiplicative risks, if \(\theta\) is low, then giving this player a reduced factor price is worse for aggregate safety than giving their competitor(s) a reduced factor price; for sufficiently high \(\theta\), subsidizing the more productive player is better.
        \item In the case where only the contest winner can cause a disaster, subsidizing the more productive player is better for aggregate safety if and only if $\theta > -1$.
    \end{enumerate}

\end{claim}

\begin{figure}
    \centering
    \includegraphics[width=0.8\textwidth]{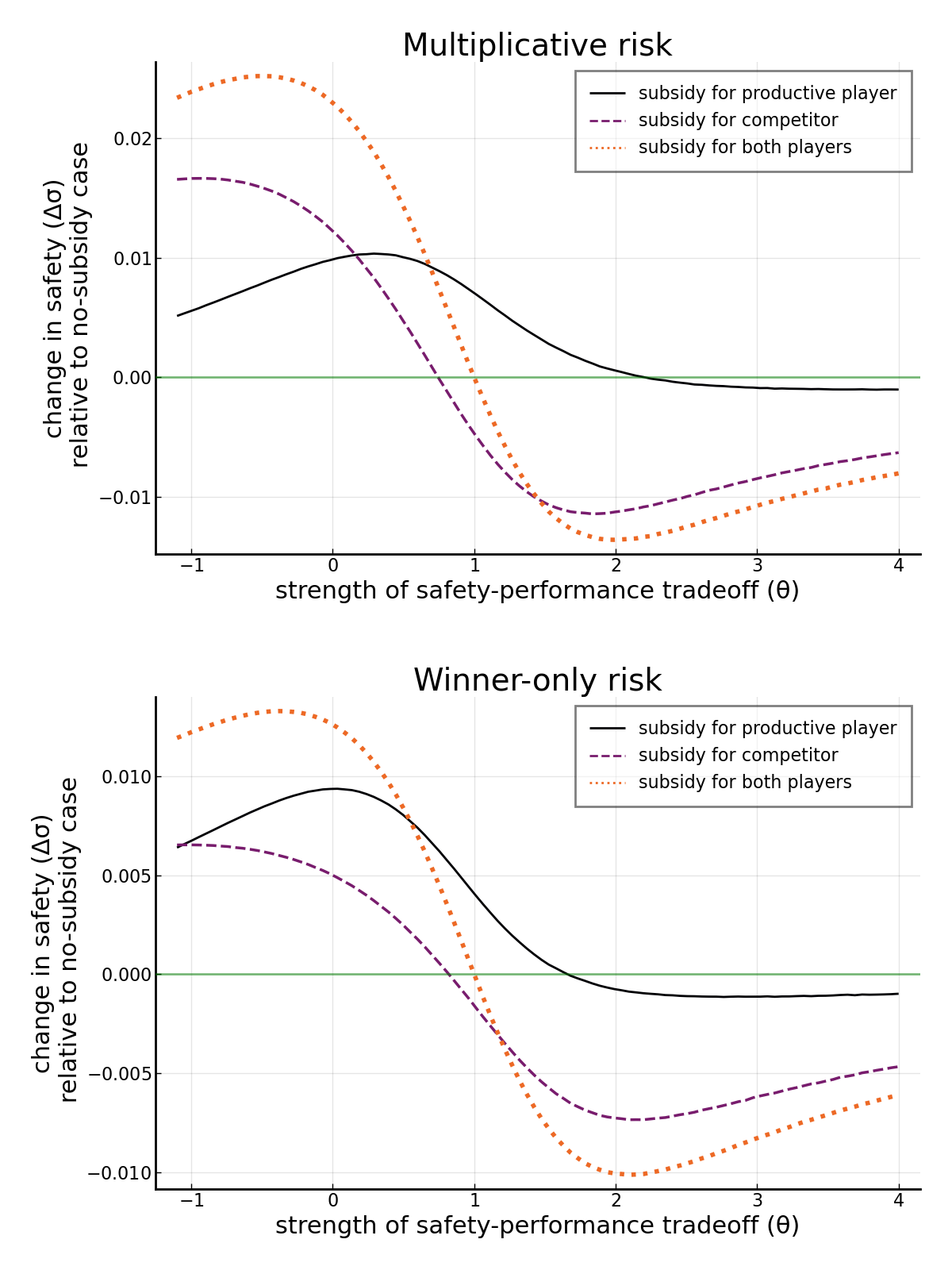}
    \caption{\textbf{Subsidies for more productive agents lead to higher safety when the safety-performance tradeoff is moderately strong.} Illustration of Claim 3, This figure shows differences in aggregate safety for various subsidy schemes, where one player has \(B\) twice as high as the other. Both the upper and lower plots use the same parameter values, with the only change being the way in which risk is aggregated. In this and all subsequent figures, the subsidized player pays \textit{half} the per-unit cost that their competitor pays.}
    \label{fig:claim-diff-B}
\end{figure}

Figure \ref{fig:claim-diff-B} illustrates this claim for both risk assumptions. On the right of the figure, where $\theta$ is high, subsidizing only the more productive player is better than subsidizing either the less productive player or subsidizing both players, but not giving out a subsidy is the best option of all. Note that the subsidy illustrated in the figure is not extremely large; if it were large enough, Claim 2 dynamics would apply. In the middle of the figure, where the safety-performance tradeoff parameter $\theta$ is high, but not too high, subsidizing only the most productive player is the best option. On the left, where $\theta$ is low, things are more complicated. The optimal policy depends on just how low $\theta$ is and how risk is aggregated, among other factors.

Lowering the price of inputs for one player has both a direct effect in changing their optimal portfolio of performance and safety investments and an indirect effect in altering the strategic scenario. When $\theta$ is low, both players are willing to invest in safety. However, giving a subsidy to the more productive player increases the capability gap between the players, forcing the less capable agent to cut corners on safety. Here, the strategic effect is relatively important: a subsidy that increases the gap between the agents is less beneficial than one that reduces it. On the other hand, when $\theta$ is high, both agents are reluctant to invest highly in safety, so by increasing the capabilities gap between agents, we take some competitive pressure off of the productive player, making them more willing to spend on safety relative to performance. Subsidizing the less productive player is less beneficial because the less productive player still has a strong incentive to cut corners on safety, driven both by her low $B_i$ and high $\theta$.\footnote{It's important to note that Claim 2 still holds here: given a large enough subsidy for either player, we can achieve high aggregate safety. This claim is relevant when we cannot provide an arbitrarily generous subsidy and want to decide whom (if anyone) to subsidize.}

We now turn to an analysis of subsidizing more and less safety-conscious players. Claim 4 examines the case where there is a large difference between how difficult the players believe it is to achieve safe outcomes. Here again, whether risk derives from the race winner, or from both competitors, influences the dynamics.

\begin{claim}\label{claim:safety-conscious}
    Suppose that only the contest winner can cause a disaster and that players differ in their beliefs about the safety productivity factor $A$. Regardless of all other parameter values, if the difference in players' beliefs about $A$ is great enough (so one player believes $A$ to be sufficiently large relative to the other), it is better for aggregate safety to subsidize the player who believes $A$ to be lower. This is not true when disaster risk is aggregated multiplicatively.
\end{claim}

\begin{figure}
    \centering
    \includegraphics[width=0.9\textwidth]{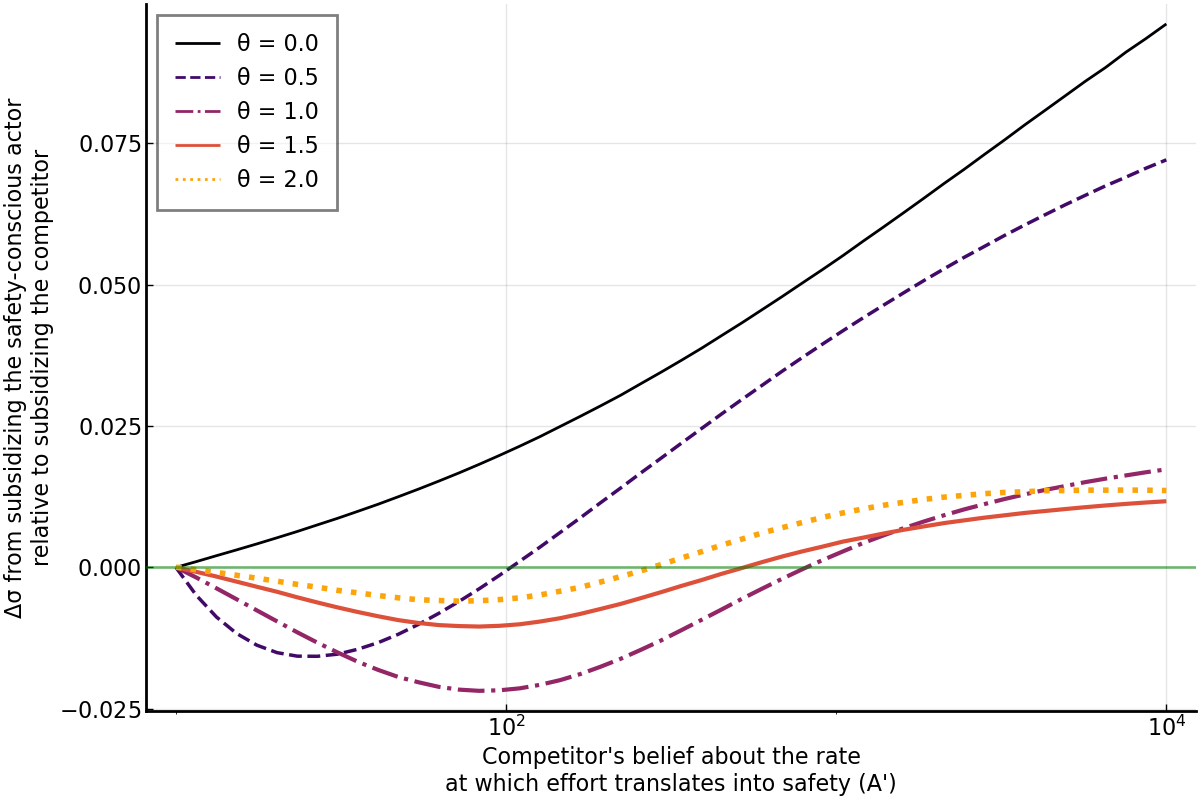}
    \caption{\textbf{Subsidies for safety-conscious agents increase safety if competitors are sufficiently unconcerned about safety}. Player 1 believes that $A = 10$, which is the true value, while player 2 believes (incorrectly) that $A = A' > 10$. Here, $\Delta \sigma$, the difference in safety for giving player 1 a subsidy rather than giving player 2 the same subsidy, is shown for a range of values of $A'$ and various values of $\theta$. We assume that only the contest winner can cause a disaster. As asserted in Claim 4, we can see that for sufficiently high $A'$, $\Delta \sigma > 0$ in all cases.}
    \label{fig:claim-A-beliefs-diff}
\end{figure}

Figure \ref{fig:claim-A-beliefs-diff} shows an example of a scenario that illustrates this claim. In this scenario, we have two players who are identical except for their beliefs about the $A$ parameter: both players have the same $A$, but they disagree about the true value of this $A$, with player 1 believing that $A = 10$, and player 2 believing that $A = A'$, which we let vary along the x-axis of the figure. On the y-axis, we measure the difference in aggregate safety from giving a subsidy to player 1, relative to giving the same subsidy to player 2. We can see that, as player 2's belief about the ease of achieving a safe outcome, $A'$, increases, it may be safer to subsidize player 2, but for very high values of $A'$, it is always better to subsidize player 1. Intuitively, if we want to promote safety, we shouldn't subsidize a player who thinks that being safe is trivially easy (i.e., a player who believes $A$ to be very high).

This claim gives us some idea of when it may be safety-promoting to assist a safety-conscious agent: if one agent thinks that being safe is trivial, it's probably best to assist that agent's competitors; however, if agents are more similar in their attitudes toward disaster risk, the question of whom (if anyone) to subsidize is more unclear.

We should note that this claim addresses only one notion of what it means to be safety-conscious; namely, we say that an agent is safety-conscious if she believes $A$ to be low relative to her competitors. In Appendix \ref{sec:appendix-diff-d}, we consider a different notion of safety-consciousness based on the cost $d$ that players face in the event of a disaster. Importantly, when we measure safety-consciousness based on players' appraisals of the costs of disaster, we find that, under many parameter values, it is better to subsidize the less safety-conscious agent, as the safety-conscious agent will not compromise as much on safety in order to compete on performance.

It is also worth noting that this claim deals with cases where all agents intrinsically value performance, and value safety only insofar as it guarantees the returns to their performance. An agent who cares purely about safety would not be subject to the same consideration.

\section{Conclusion}

In this paper, we develop and solve a simple baseline model for competitive AI development to provide policy recommendations for third parties concerned about risks from competition to develop a new technology. We demonstrate that the effect on safety of lowering the price of inputs to AI developers depends on whether performance or safety scales more rapidly with compute. From our model, we see that there are a number of potential scenarios in which lowering the price of inputs leads to increased safety. The first case is when safety scales more rapidly with increases in safety research than with reductions in performance research. For example, \cite{bai_training_2022} find that both helpful and harmless language models scale similarly with the number of parameters in the model.\footnote{Though this is not compute scaling, the two are positively correlated (see e.g. \cite{sevilla_estimating_2022}).} The second case is if the principal is able to price discriminate, offering different prices to each player: she can increase safety by giving one player a subsidy large enough to discourage risky competition from that player's competitors.

Next, we examine the case of heterogeneous agents. We find that, when there is a steep safety-performance tradeoff, it is better to subsidize the agent who is more productive at performance research. Finally, depending on how risk is aggregated between players, if one player is much more safety-conscious than others, it is better for safety to subsidize that player rather than to subsidize any of her competitors.

Although subsidizing a much more safety-conscious, or productive, agent often improves safety as intuition suggests, subsidizing a somewhat more safety-conscious or productive agent can often be harmful. Very large subsidies to one agent are beneficial when smaller subsidies are not. We should not allow intuitions derived from extreme cases to govern considerations of more moderate interventions and cases.

We hope that this model provides a foundation that can be built on in future work. One promising line of research would be to explore ways the principal could contract with agents to incentivize them to commit to a certain level of safety in return for a compute discount, potentially altering the range of scenarios over which agents would agree to reduce competition in exchange for resources (\cite{stafford_IAEA_2022}). A second line of research might examine the role of information in optimal compute provision, studying the cases in which sharing productivity-enhancing insights between agents is safety-promoting, and understanding better the welfare loss that results if agents are able to conceal their true preferences from the principal. Finally, one could adapt the model presented here to allow for explicit consideration of agents' strategies over multiple time periods, which could enable us to better model accrual of technology (via investment or information diffusion) and more faithfully represent general racing dynamics over time.

\bibliographystyle{plainnat}
\bibliography{main}

\newpage

\appendix

\begin{center}
    \Large{Appendix}
\end{center}

\section{Description of the numerical solver}\label{sec:appendix-solver-description}

The numerical solver works by iterating on players' best responses to each other's strategies. Assuming that we have specified $u_i$ for each player $i$, we propose some strategy $X_{s,i}^{(0)}, X_{p,i}^{(0)}$ for each $i$ and proceed iteratively:

On each iteration $t$, we update each $X_{s,i}^{(t)}, X_{p,i}^{(t)}$ to the values that maximize $u_i$ given the other players' strategies from the previous iteration:
\begin{equation}
    X_{s,i}^{(t)}, X_{p,i}^{(t)} = \argmax_{X_{s,i}, X_{p,i}} \left\{ u_i(X_s, X_p) \right\}
\end{equation}
$$\text{s.t. } X_{s, -i}, X_{p, -i} = X_{s, -i}^{(t-1)}, X_{p, -i}^{(t-1)}$$

We continue iterating until the proposed strategies stop changing, within some error threshold, between iterations. Assuming a negligible error threshold, the strategies we end with will be a Nash equilibrium, since by construction no player has an incentive to change their strategy. Although the iteration is not guaranteed to converge, this method works quite well in practice.

The code used to represent and numerically solve the model (with the above-described algorithm) is available \href{http://github.com/quevivasbien/AIIncentives.jl}{on GitHub}.

\section{Procedure for numerical ``proofs'' of claims}

Due to the complexity of our model, it is difficult to analytically prove more than trivial claims about the model's behavior. Therefore, we rely on the results of our numerical solver to generate and verify the claims presented in this paper. The procedure we use can be described as follows:

\begin{enumerate}
    \item We start with a hypothesis $H$, which specifies that if parameters are selected from some set $\Theta_H$, then a proposition $P$ must hold.
    \item We choose some finite subset $\hat \Theta_H \subseteq \Theta_H$ of test points. Ideally, this subset should be representative of the entire relevant parameter space; in testing the claims, we have selected reasonably exhaustive test points with a focus on intuitively likely regions of the parameter space, though this is of course subject to computational constraints and our judgments about which parts of the parameter space merit the most scrutiny.
    \item For each $\theta \in \hat \Theta_H$, we solve for the equilibria of the problem parameterized by $\theta$, and check that $P$ holds.
    \item If $P$ holds for all $\theta \in \hat \Theta_H$, we accept $H$ as a claim.
\end{enumerate}

Because we verify claims by testing them at finitely many points, these claims are of course not strict mathematical results but can be regarded as robust observations about the model's behavior.

\section{Behavior when players differ in $d$}\label{sec:appendix-diff-d}

Suppose that one player believes that a disaster will be more costly
(has a higher \(d\)) relative to their competitor(s), with players
otherwise identical. It is difficult to describe the circumstances under which a subsidy for one player is better for safety than the same subsidy for that player's competitor, but we can give some general rules here:

\begin{itemize}
    \item In the case of multiplicative risks, if \(\theta\) is low, giving this player a reduced
    factor price is worse for aggregate safety than giving their
    competitor(s) a reduced factor price; for higher \(\theta\), the
    situation is ambiguous, with subsidies for the player who believes
    disasters to be more costly being better for aggregate safety only if some
    combination of the following holds:
    
    \begin{itemize}
        \item \(A\) is sufficiently high
        \item \(\alpha\) is sufficiently low
        \item \(B\) is sufficiently low
        \item \(\beta\) is sufficiently high
        \item \(r\) (unsubsidized) is sufficiently high
    \end{itemize}
    
    \item In the case where only the contest winner can cause a disaster, the above holds with the caveat that a subsidy for the player who believes a disaster is more costly is also better for aggregate safety (relative to a subsidy for their competitor) if $\theta$ is near zero and $\alpha$ and $\beta$ are sufficiently low and high, respectively.
\end{itemize}

\begin{figure}
    \centering
    \includegraphics[width=0.8\textwidth]{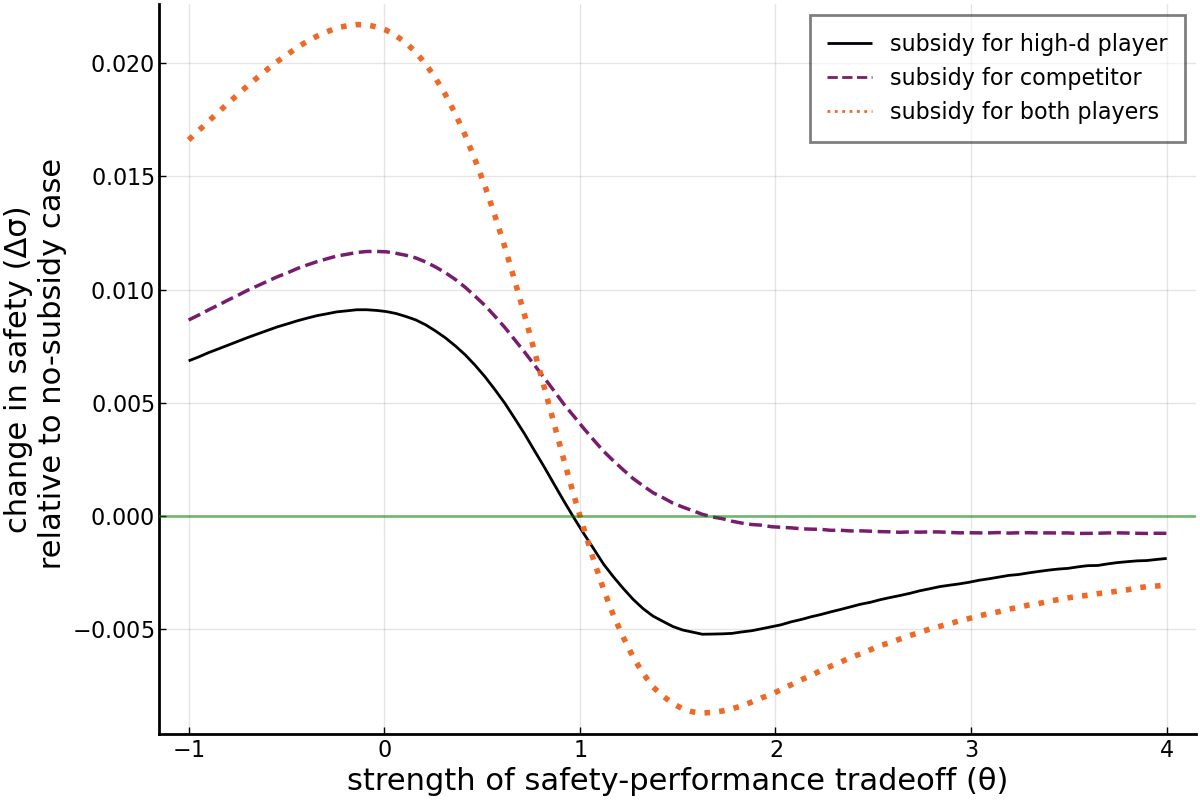}
    \caption{Example showing differences in aggregate safety for various subsidy schemes. One player has \(d\) twice as high as the other (meaning they believe disasters to be more costly than the other does). Here, subsidies for the low-\(d\) player result in higher safety than subsidies for the high-\(d\) player at all levels of \(\theta\).}
    \label{fig:claim-d-diff-a}
\end{figure}

\begin{figure}
    \centering
    \includegraphics[width=0.8\textwidth]{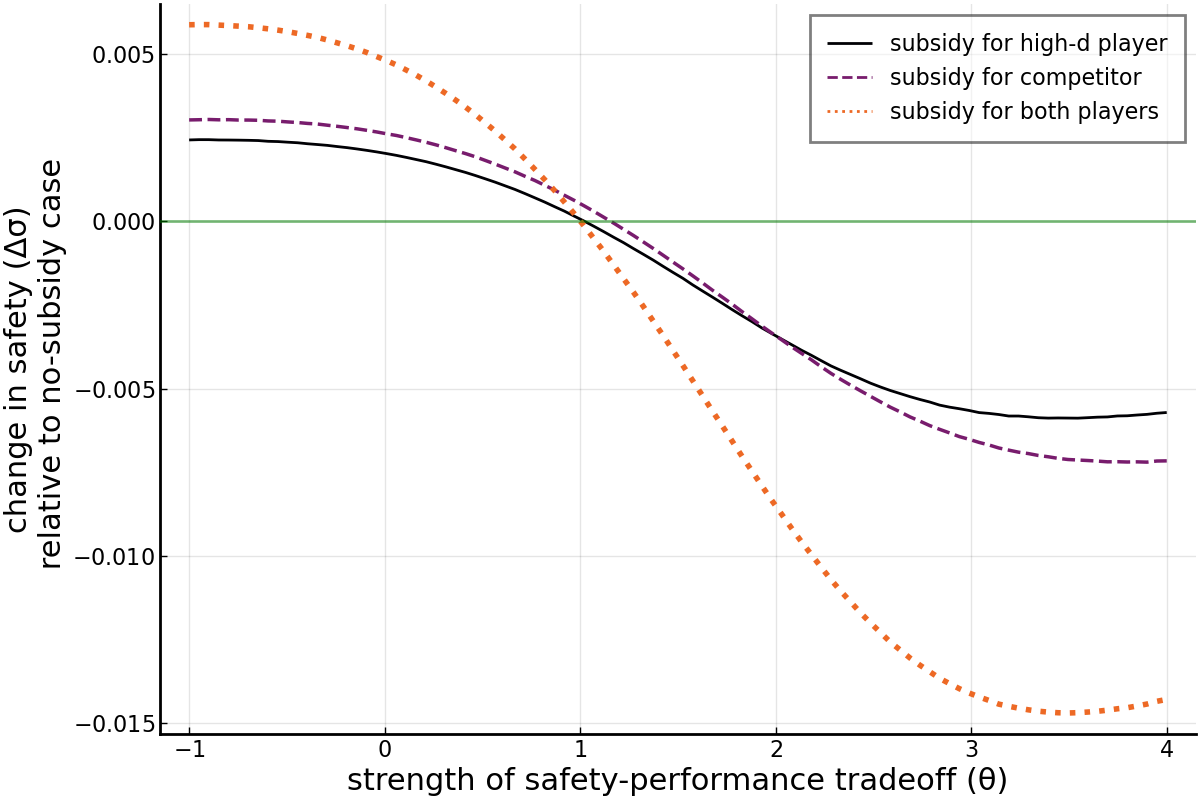}
    \caption{Same as Figure \ref{fig:claim-d-diff-a}, but with \(A\) 10x and \(B\) 0.5x values in Figure \ref{fig:claim-d-diff-a}. Here, subsidies for the player who believes disasters to be more costly (high \(d\)) result in higher safety for sufficiently high \(\theta\).}
    \label{fig:claim-d-diff-b}
\end{figure}

\begin{figure}
    \centering
    \includegraphics[width=0.8\textwidth]{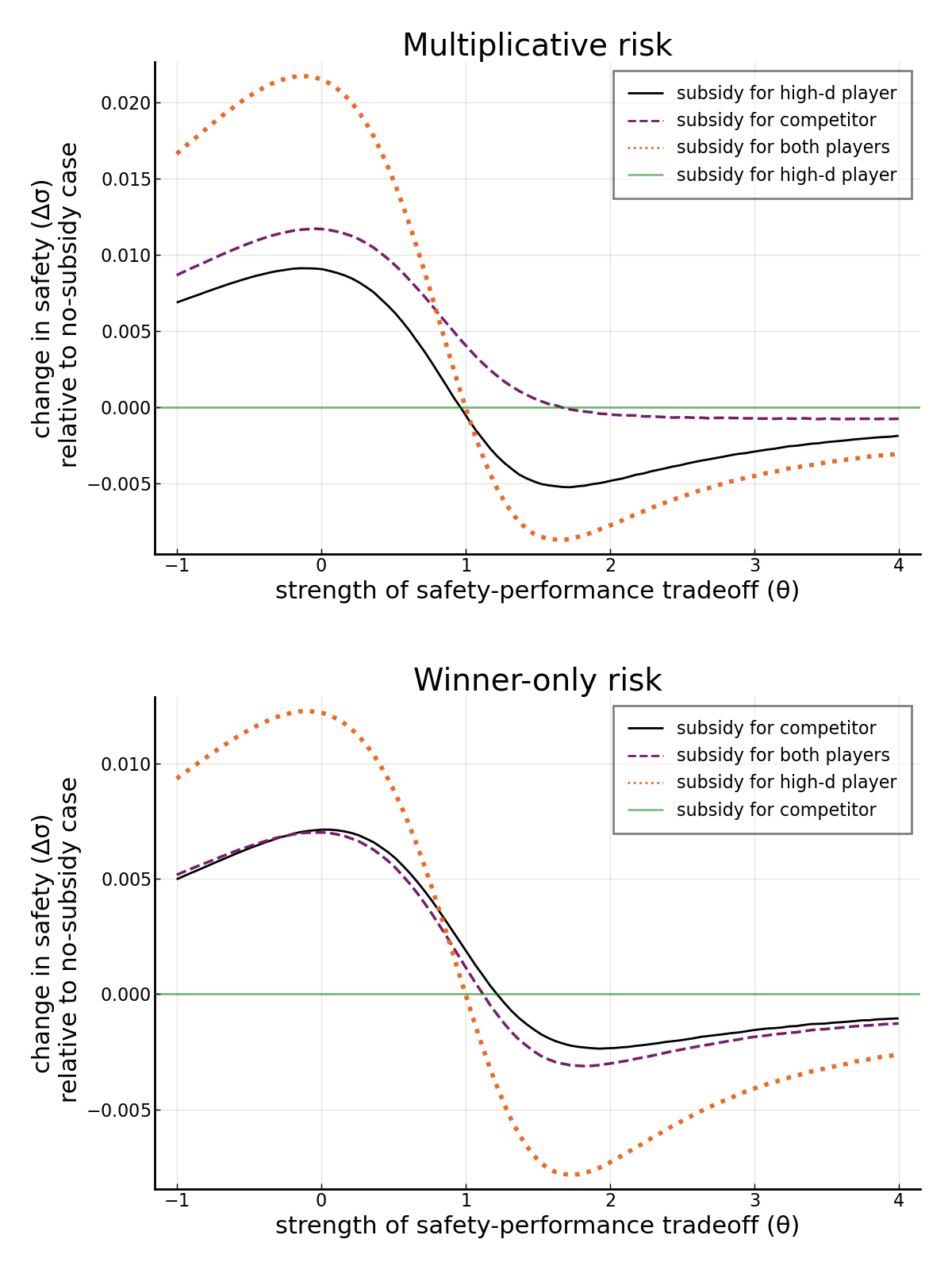}
    \caption{Example comparing risk assumptions, where one player has \(d\) twice as high as the other (meaning they believe disasters to be more costly that the other does). Parameter values used are the same as Figure \ref{fig:claim-d-diff-a}.}
    \label{fig:claim-d-diff-c}
\end{figure}

Some scenarios illustrating this behavior are shown in Figures \ref{fig:claim-d-diff-a}, \ref{fig:claim-d-diff-b}, and \ref{fig:claim-d-diff-c}. In Figure \ref{fig:claim-d-diff-a}, a subsidy for the player who faces a lower cost from disaster (lower $d$) is better for safety at all levels of $\theta$; Figure \ref{fig:claim-d-diff-b} shows a different result, where a subsidy for the player with higher $d$ is better at high levels of $\theta$. We assume that risk is aggregated multiplicatively in both of these figures.

Figure \ref{fig:claim-d-diff-c} compares two scenarios with the same parameter assumptions but different assumptions about risk aggregation. In this example, a subsidy for the player who faces a lower cost from disaster (lower $d$) is better for safety at all levels of $\theta$ in the case of multiplicative risk aggregation but only for modestly high values of $\theta$ in the case where only the contest winner can cause a disaster.

What is driving this counterintuitive result? Intuitively, the risky player is more susceptible to competitive pressure to increase performance and will tend to respond to price changes by favoring performance over safety more than the safety-conscious player would. Granting a subsidy to the safety-conscious agent causes the other agent to cut corners on safety to remain competitive, and the safety-conscious agent may not be able to sufficiently compensate for this by increasing their own safety spending. On the other hand, subsidizing the risky player causes her to have an advantage in the race, reducing the pressure to spend on performance and thus allowing for more spending on safety; the unsubsidized safety-conscious player will not sacrifice as much on safety in order to compete. The overall effect may be that subsidizing the risky agent is more beneficial.

\end{document}